\documentstyle[prl,aps]{revtex}
\input epsf

\tighten

\begin{document}


\newcommand{\Bd}{{\dot B}}
\newcommand{\Cd}{{\dot C}}
\newcommand{\fd}{{\dot f}}
\newcommand{\hd}{{\dot h}}
\newcommand{\ep}{\epsilon}
\newcommand{\vp}{\varphi}
\newcommand{\al}{\alpha}
\newcommand{\be}{\begin{equation}}
\newcommand{\ee}{\end{equation}}
\newcommand{\bea}{\begin{eqnarray}}
\newcommand{\eea}{\end{eqnarray}}
\def\gapp{\mathrel{\raise.3ex\hbox{$>$}\mkern-14mu
              \lower0.6ex\hbox{$\sim$}}}
\def\gsim{\gapp}
\def\lapp{\mathrel{\raise.3ex\hbox{$<$}\mkern-14mu
              \lower0.6ex\hbox{$\sim$}}}
\def\lsim{\lapp}
\newcommand{\PSbox}[3]{\mbox{\rule{0in}{#3}\includegraphics{#1}\hspace{#2}}}
\def\Tr{\mathop{\rm Tr}\nolimits}
\def\su#1{{\rm SU}(#1)}

\title{
Observation of Cosmic Acceleration and Determining the Fate
of the Universe}

\author{
Glenn Starkman, 
Mark Trodden and 
Tanmay Vachaspati}
\address
{Department of Physics,
Case Western Reserve University,
10900 Euclid Avenue,
Cleveland, OH 44106-7079, USA.}

\wideabs{
\maketitle

\begin{abstract}
\widetext
Current observations of Type Ia supernovae provide 
evidence for cosmic acceleration out to a redshift of $z \lsim 1$,
leading to the possibility that the universe is entering an
inflationary epoch. However, inflation can take place
only if vacuum-energy (or other sufficiently slowly redshifting source 
of energy density) dominates the energy density of a region of physical 
radius $1/H$. We argue that for the 
best-fit values of $\Omega_\Lambda$ and $\Omega_m$ inferred from 
the supernovae and other data, 
one must confirm cosmic acceleration out
to at least $z \simeq 1.8$ to infer that our portion of the universe is 
in principle capable of inflating;
but {\em no measurement} will be able to confirm 
or falsify that inference until $\Omega_\Lambda$ rises to $0.96$.

\end{abstract}
}

\narrowtext



Recent direct \cite{perl,kirsh} 
measurements of the cosmic expansion using supernova at redshifts
from $z=0-1.2$, suggest that the expansion of 
our portion of the universe is { \it accelerating}.
This supports earlier indirect evidence leading to 
the same conclusion \cite{kraussturner,ostriker,krauss}.
This is unlike the deceleration expected in a universe 
dominated by the energy density of ordinary or dark matter.
It implies that the expansion is driven by 
the energy density of the vacuum.
This vacuum energy has variously been ascribed to
a cosmological constant in the Einstein equations,
to the zero-point fluctuations of quantum fields,
so-called vacuum energy, 
to the potential energy density of dynamical fields, 
quintessence, and to a network of cosmic strings.

In exploring the implications of these explanations,
it has widely been assumed that the energy density driving
the accelerating expansion is homogeneous.
If so, then unless the field dynamics are chosen to avoid it, 
the universe has entered on an extended period of rapid growth --
a new epoch of inflation  -- in which objects currently
within our observable universe will soon begin to leave it.
If correct, this result could have dramatic implications
for our understanding of fundamental processes underlying the Big Bang.
However, one must be careful to separate the observational  results 
from the assumptions which are built  into the  standard
interpretations of them.  

The  observational situation can be briefly summarized:  
measurement of the light curves of several tens of Type  Ia supernovae 
have allowed accurate measurements of 
the scale factor as a function of redshift ($a(z)$) out to  $z\lsim1$.  
When the supernova data in conjunction with CMBR anisotropy
data are fit to a family of cosmological models parametrized
by a homogeneous vacuum energy density  
and a homogeneous matter energy density 
(characterized by the ratios $\Omega_\Lambda$ and $\Omega_m$
of these  energy densities to the critical energy density 
$\rho_c = 3H_0^2/8\pi{G}$, all at the present epoch),
then the measured functional form of $a$
is most consistent with $\Omega_\Lambda\simeq 0.8$ and $\Omega_m\simeq 0.2$.
Notwithstanding any debates about systematic uncertainties,
for the purposes of this paper let us accept this data  as reported.
Still, the data do not imply that the observable universe 
out to the last scattering surface and beyond is vacuum-energy dominated.
We could be living in a bubble of high vacuum energy density 
surrounded by  much lower, even zero, vacuum energy  density.  
For example, the vacuum energy density could be due to a scalar field 
which locally deviates from the minimum of its potential.

The question now arises -- 
will the local vacuum dominated region inflate or will it not?
The answer to that question depends on what
one means by ``inflate''.  
The effective scale factor of our local corner of the universe 
{\it is} apparently experiencing a period of accelerated growth.
However, the essence of inflation is not local acceleration,
but acceleration over a large enough region to affect the causal
relationships between (comoving) observers.   
In particular, if the acceleration is taking place 
over a sufficiently large region, 
then unless the acceleration is halted,
comoving observers from whom one was previously able to receive signals,
will disappear from view\cite{lmkgds}. To be precise, 
much as if we were watching them fall through the horizon of a black hole,
we will see them freeze  into apparent immobility,
the signals from them declining indefinitely in brightness and energy.

The question we address in the balance of this letter
is how far out must one look to infer that the patch
of the universe in which we live is inflating?


In any Friedman-Robertson-Walker
(FRW) cosmology, there exists for each comoving observer
a sphere centered on that observer, on which the velocity of
comoving objects is the speed of light.   
When sources inside that sphere emit radially inward-directed light-rays, 
the photons approach the observer;
when sources outside that sphere emit radially inward-directed light-rays, 
the physical distance between the photons and the observer increases.
This sphere is  the minimal anti-trapped surface (MAS).
For a homogeneous universe, the physical radius of the MAS is $1/H$, 
since the physical velocity $v$ of a comoving observer 
at a physical distance $x$ is $v=cHx$.  

In a matter or radiation dominated epoch, 
when the dominant energy density in the universe 
scales with the scale factor $a$ as
$\rho_{dom}\propto a^{-n}$ with $n>2$, 
the comoving radius of the MAS grows -- 
the inward-directed photons outside the MAS,
which were making negative progress in their journey to the observer, 
eventually find themselves inside the  MAS, and reach the observer.
New objects are therefore constantly coming into view.
If, on the other hand, the dominant energy density in the universe 
scales as the scale factor to a power greater than $-2$
then the comoving radius of the MAS is contracting.
For an equation of state with 
$\lim_{t\to\infty}n < 2$,
the MAS contracts to zero comoving radius in finite conformal time, 
$\eta_{max}$, corresponding to time-like infinity.
The interior of the past null cone  of the observer at $\eta_{max}$ 
is the entire portion of the history of the universe that the observer can see,
what we shall call hereafter the observer's 
visible history of the universe (VHU). 
Since the  null cone is contracting, 
comoving objects cross out of the null cone,
and thus disappear from view (see Fig. 1).
More precisely, the  history of the objects 
after the time when  they cross out of the null cone is unobservable.
Again, much as when watching something fall through the horizon of a black hole,
the observer continues to receive photons from the disappearing source
until $\eta_{max}$, but the source appears progressively redder and dimmer,
and its time evolution freezes at the moment of horizon-crossing.

The comoving contraction of our MAS, and particularly 
the  motion of comoving sources out of our VHU
is the essence of inflation.
These sources can never again be seen, unless the 
equation of state changes so that $n>2$
and the MAS grows once more, i.e. inflation ends.
(Then, the sources never actually crossed out of the VHU,
merely out of the apparent VHU -- the VHU
one would have inferred without the change in equation of state.)

By examining the Raychaudhuri equation governing the
evolution of the divergence of geodesics, 
Vachaspati and Trodden \cite{VacTro98} recently showed that at 
any time $\eta_e$ a contracting anti-trapped surface 
cannot  exist in the universe unless a region of  
radius greater than $1/H(\eta_e)$ is vacuum dominated and homogeneous.
(The result depends on the validity of the weak energy and certain 
other conditions. In the present application, the conditions
seem reasonable.) Thus no inflation will have occurred unless a 
region of size $1/H(\eta )$ remains vacuum dominated long enough
for the  MAS to begin collapsing.
We apply this bound to determine our
ability to infer the current and future state of our
patch of the universe.


The geometry of a homogeneous and isotropic universe is described
by the FRW metric:
\begin{equation}
d s^2 = a^2 \left[d\eta^2 - {d r^2\over1-kr^2} 
- r^2\left(d\theta^2+\sin^2\theta d\phi^2\right)\right]
\end{equation}
$a(\eta)$ encodes the changing relationship
between coordinate (comoving) distances and physical distance.
(The conformal time $\eta$ is related to the proper time $t$ 
measured by comoving observers by $a(\eta) dt = d\eta$.)
The evolution of $a$ is determined by
the mean energy density of the universe $\rho$
and by the curvature radius of the geometry, characterized
by $k$, via the Friedmann equation:
\begin{equation}
\label{eqn:friedmann}
\left( {{\dot a}\over a^2}\right)^2 = {8\pi G \over 3}\rho - {k\over a^2}  ,
\end{equation}
where ${\dot a}\equiv(da/d\eta)$, 
and $({\dot a}/a^2)\equiv H$ is the Hubble parameter; 
currently $H \equiv H_0 = 72$ km/s/Mpc
\cite{hubbleparam}.

\begin{figure}[tbp]
\caption{\label{figure1}
Spacetime diagram showing the MAS position ($r=1/aH$) as a function of
conformal time $\eta$ where the universe eventually does (solid line)
and does not (dashed line) undergo inflation.
On our present lightcone we observe supernovae
(SN) and the cosmic microwave background (CMB).
If the universe inflates, the MAS curve turns around at $\eta_c$
and the interior of the past light-cone of
an observer at future infinity $\eta_{max}$ only covers a portion
of the spacetime (triangular area under the dashed line marked
VHU). In this case, the first object to leave causal contact with
us will do so at $\eta_d$ through the point P.
}
\epsfxsize = 0.85 \hsize \epsfbox{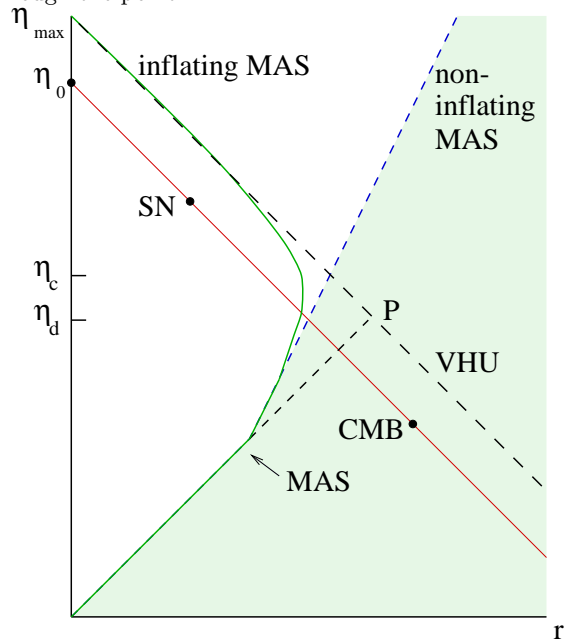}
\end{figure}

\vskip -0.1 truecm
The energy density $\rho$ can receive many distinct contributions,
but what is crucial to the nature of the solution to (\ref{eqn:friedmann})
is the fraction of the critical density $\rho_c= 3 H_0^2/8\pi G$
comprised by each species, and how each of these contributions scales 
with $a$. The two most important possible contributions today, 
non-relativistic matter and vacuum energy, scale respectively as $a^{-3}$
and $a^0$. 

Since the metric (\ref{eqn:friedmann}) describes a homogeneous and 
isotropic space, we can choose, without loss of generality,
to locate ourselves at the origin of coordinates, $r=0$.
Consider then a source located at a  comoving distance $r$ from us.
At conformal time $\eta_e$ the source emits a photon  directed toward us;
the photon is received, and the source therefore observed, 
at $r=0$ at our present conformal time $\eta_0$.
If the source is a standard candle of known luminosity ${\cal L}$,
then by measuring the observed flux ${\cal F}$ of the source,
one can determine its luminosity distance:
\begin{equation}
d_L = \left({{\cal L} / 4 \pi {\cal F}}\right)^{1\over 2}
\end{equation}
If one can measure the redshift $z$ of the source,
then one can 
constrain $\Omega_M$ and $\Omega_\Lambda$ using the relationship
\begin{eqnarray}
d_L(z) = c H_o^{-1} (1+z) \vert\Omega_k\vert^{-1/2}
{\mathrm sinn}\bigl[\vert\Omega_k\vert^{1/2}&\cr
\int_0^z dz'\bigl[(1+z')^2(1+\Omega_M z') 
- z'(2+z')&\Omega_\Lambda\bigr]^{-1/2}\bigr] .
\end{eqnarray}
Here $\Omega_k\equiv1-\Omega_M-\Omega_\Lambda$ 
and  sinn$(x)$ is $\sin(x)$ if $k>0$ and $\sinh(x)$ if $k<0$.
Repeating this for a variety of sources at different comoving
distances allows one to determine $\Omega_M$ and $\Omega_\Lambda$.
In Fig. 2, we plot the difference in apparent magnitude
between a flat $(k=0)$ homogeneous vacuum-energy dominated cosmology
($\Omega_\Lambda=0.4,0.6,0.8$)  
and the best fit vacuum-energy free cosmology, 
a negatively-curved ($k<0$) universe  with $\Omega=0.3$.

\begin{figure}[tbp]
\caption{\label{figure2}
The difference in apparent magnitude between a flat FRW universe
($\Omega_\Lambda+\Omega_{\rm matter}=1$)
for $\Omega_\Lambda =$0.4 (dashed curve), 0.6 (dotted curve)
and 0.8 (solid curve), and an $\Omega_\Lambda =0$,
negatively curved universe with $\Omega=0.3$,
{\it vs.} the redshift $z$.
The dashed-dotted curve shows $z_{MAS}$ for the different values of
$\Omega_\Lambda$.
}
\vskip -0.3 truecm
\epsfxsize = \hsize \epsfbox{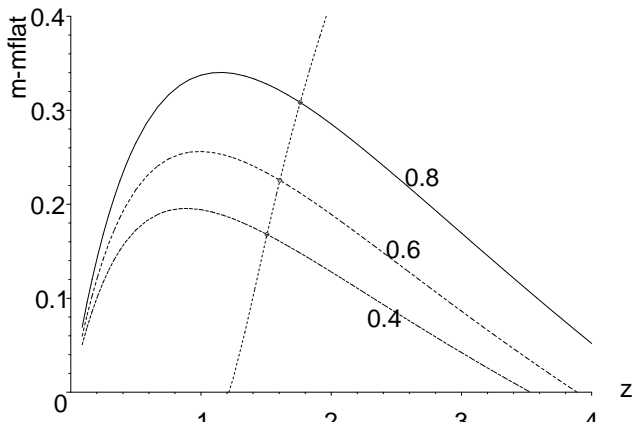}
\vskip -0.6 truecm
\end{figure}

This approach has recently been applied to Type Ia supernovae
by two independent groups \cite{perl,kirsh}.  
Looking at tens of  supernovae out to a redshift of $z<0.83$,
both groups find that the best-fit values of $\Omega_\Lambda$ and 
$\Omega_m$ are approximately $0.8$ and $0.2$ respectively.


The physical distance between the emitting supernova and the observer
at time the time of emission  $\eta_e$ was
\begin{equation}
d(\eta_e) = a(\eta_e) \int_{\eta_e}^{\eta_0}{d\eta}  .
\end{equation}
where, for simplicity, we have now restricted our attention
to the case when the geometry of the universe is flat ($k=0$). 
We are seeing our MAS if 
$d(\eta_e)=1/H(\eta_e)$, i.e.
\be
{\sqrt{1 + \alpha (1+z)^3}\over  (1+z)} 
\int_1^{1+z} {{dy}\over \sqrt{1+\alpha y^3}} = 1,
\label{masol}
\ee
where $\alpha \equiv (\Omega_\Lambda^{-1}-1)$.
In Fig. 3, we plot $z_{MAS}$, the redshift of the MAS,
{\it vs.} $\Omega_\Lambda$.
(In Fig. 2, the dash-dotted line shows both the redshift 
$z_{MAS}$ and the associated magnitude difference
from the best fit negatively-curved universe.)
We find that for the best fit  value ${\Omega_\Lambda=0.8}$,
we will see the MAS if we look out to $z=1.8$.
If observations continue to find cosmic acceleration 
with $\Omega_\Lambda=0.8$ out to $z=1.8$, 
then we can infer that our MAS is contracting,
{\it i.e.}~ our patch of the universe is inflating,

Fig. 3 shows the redshifts out to which one must
necessarily ascertain cosmic acceleration for any given value
of vacuum energy before one can be confident that inflation is a
possible fate of the universe.

\begin{figure}[tbp]
\caption{\label{figure3}
$z_{MAS}$ versus $\Omega_\Lambda$. The curve specifies the
redshift out to which it is necessary to confirm cosmic acceleration
so that it might lead to an inflationary universe.
}
\vskip -0.3 truecm
\epsfxsize = \hsize \epsfbox{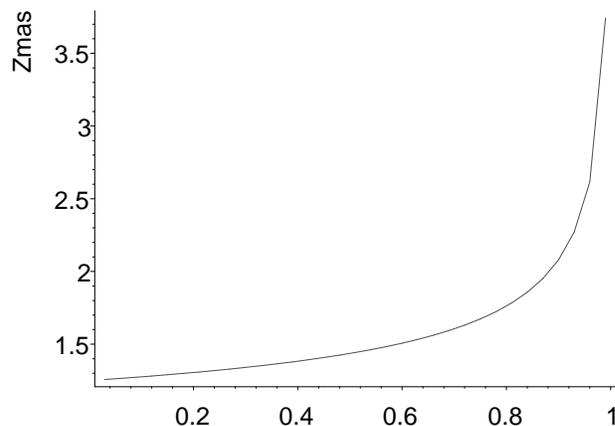}
\vskip -0.6 truecm
\end{figure}

What if observations find a precipitous decrease in $\Omega_\Lambda$
between $z=0.8$ and $z=1.8$?  Because we are looking along the 
past null cone, this decrease could be due either to a negative spatial
gradient in  the vacuum energy $\rho_{vac}$, or to a temporally increasing
$\rho_{vac}$. In case  of a gradient, we would conclude
that  we are in a vacuum bubble and the universe will not inflate,
at least not with $\Omega_\Lambda=0.8$.
However, we cannot exclude the possibility that the vacuum energy 
was growing with time.
It is not easy to get an increasing $\rho_{vac}$ -- 
in the long time limit fields tend to evolve to the minima of 
their effective potentials.
For $\rho_{vac}$ to have increased dramatically
between $z=1.8$ and $z=0.8$, requires the responsible field to have
had sufficient kinetic energy to climb up its potential.  
But field kinetic energies scale rapidly with scale factor
(as $a^{-6}$). If the kinetic energy was not to have dominated
the universe at redshifts greater than 5,  that kinetic energy
had to have been generated by a recent conversion of potential energy. 
Thus the ``inflaton'' must have rolled down one side
of a potential well and back up the other.  One cannot exclude
such a possibility, but the required potentials and the necessary
accidents of timing must be carefully tailored to fit the data.

Are the current data on the fluctuations
of the cosmic microwave background temperature  consistent 
with us living inside a vacuum bubble?  The answer is likely 
model dependent.  We believe the dominant bubble contributions to the 
CMBR fluctuations will come from the time evolution of the bubble wall
and the integrated Sachs-Wolfe effect,  and so depend on 
the wall velocity, energy-density profile, shape, {\it etc}.
This is a fertile subject for future investigations.

Let us return to Fig. 1 in which,
because comoving distance and conformal time are the coordinates, 
light rays propagate at $45^o$.
We see that during the matter and radiation dominated phases
the MAS grows, so that new objects at fixed comoving radius $r$
are constantly entering the interior of the MAS. 
More importantly, for the observer at $r=0$, 
new objects are constantly coming into view.
As the universe becomes vacuum dominated,
the expansion of the MAS decelerates 
and it eventually begins to contract.
We can readily find the  value of the scale factor at
which that happens by setting 
\be
\label{eqn:MAScontract}
0 = {{\dot r}_{MAS}} 
{\bf \propto } {d a \over d \eta} {d\over d a}
\left[ \sqrt{8\pi G \rho_c \Omega_\Lambda \over 3} 
  \sqrt{a^2 + {1-\Omega_\Lambda\over \Omega_\Lambda}
     {a_0^3\over a}}\right] .
\ee
We have assumed $k=0$ and ignored the energy density 
in radiation.  Eq. (\ref{eqn:MAScontract}) is satisfied 
when  $a=a_c$ with
\be
\label{MAScontract}
{a_0\over a_c} = \left( {2\Omega_\Lambda\over1-\Omega_\Lambda}\right)^{1/3} .
\ee
We see that the MAS begins to contract at an epoch $\eta_c$
when $\Omega_\Lambda (\eta_c ) =1/3$. 
For a present value of $\Omega_\Lambda=0.8$ this occurs at
$a(\eta_c) = a_0/2$.

When we look out to $z_{MAS}$, will we see this contraction?  
The answer is not immediately clear, since from Fig. 1 we see that,  
well after $\eta_c$, the past null cone of the observer at $r=0$ 
intersects the observer's MAS below the turnover.
Defining $\eta_v$ ($a_v$) to be the  time
(scale factor) when  the turnover in the MAS comes into view,
we see that $\eta_v$ is given by
\be
a_c \int_{\eta_0}^{\eta_c}d\eta = H(\eta_{c})^{-1} \ .
\ee
This can easily be rewritten as
\be
\int_{a_c/a_v}^1{d x\over \sqrt{1+2 x^3}} = {1 \over \sqrt{3}}
\ee
(including only matter and vacuum energy contributions to $H$),
which  can be solved numerically to give
$a_v \simeq 4.369 a_c$.
Combining this with equation (\ref{MAScontract}),
we see that
\be
{a_v\over a_0} = 4.369 \left(1-\Omega_\Lambda\over2\Omega_\Lambda\right)^{1/3}
\ee
For $\Omega_\Lambda \geq 0.96$, $a_v\leq a_0$.
Thus only for $\Omega_\Lambda \geq 0.96$ is it possible to 
look out far enough to infer the contraction of our MAS.
Since all observations agree that $\Omega_\Lambda < 0.96$,
it is currently  impossible to make observations of
a contracting MAS.

The contraction of the MAS {\it is} in some sense the onset of inflation;
indeed we could reasonably define it as such. 
This is because it is the projected contraction of our MAS  to zero
comoving radius in finite conformal time which  is the 
cause of the eventual loss of contact with comoving observers.
We see in Fig. 1, that our VHU is an inverted cone whose
apex is at the intersection of our MAS with our worldline, $r=0$.
However, if the loss of contact with previously visible objects
is what we consider the defining characteristic of inflation, 
then this can begin either before or after $\eta_c$, depending
on the details of the transition from expanding to contracting MAS.

The value of $\eta_d$ is easily obtained, 
since $\eta_d = \eta_{max}/2$.
For $\Omega_\Lambda=0.8$, $\Omega_m=0.2$,
we find $a_0/a_d \simeq 3$. Since this is greater than $a_o/a_c$, 
if we look out to $z_{MAS}$ and see the contraction of our MAS,
then objects may have already left our apparent VHU.

As we implied above, 
while the value of $\eta_d$ may be of some philosophical interest,
there is no measurement that one can make that guarantees
that objects have left the VHU,
since that takes an infinite amount of proper time for
an observer to observe.  
If inflation is driven by the metastable
or unstable potential energy of some quantum field,
then the MAS could eventually start expanding once again
and objects which were almost out of view, 
could come back into view.


In conclusion, if present observations of cosmic acceleration
with $\Omega_\Lambda$ (suggested to be $\simeq 0.7-0.8$)
do not extend to a redshift of $z_{MAS} (\Omega_\Lambda )$
(equal to 1.8 for $\Omega_\Lambda = 0.8$),
then either we live in a sub-critical vacuum bubble which cannot, 
on its own, support inflation,
or fine-tuned field dynamics led to a rare period of growth
in the vacuum energy immediately preceding its domination
of the energy density.
If, on the other hand, future observations 
confirm the acceleration up to and beyond a redshift of 1.8, 
this still will not show a contracting MAS, because the
accelerated expansion is not yet sufficiently advanced
for the MAS to contract --- the  universe is not yet inflating.
Either way we will have discovered exotic fundamental physics;
however, we will never be able to tell for certain if objects are
moving out of our causal horizon - because that would 
require observations out to infinite redshift. 

We are grateful to L. Krauss for important discussions,
and to D. Olson for pointing out an error in equation
(10) of an earlier version of this paper. 
This work was supported by the Department of Energy (D.O.E.)
and NSF CAREER grant to GDS.

\end{document}